\documentclass[aps,nofootinbib,prd,class-pre]{revtex4} 

\newcommand{\be}{\begin{equation}}\newcommand{\ee}{\end{equation}}
\newcommand{\bea}{\begin{eqnarray} }\newcommand{\eea}{\end{eqnarray}}
\newcommand{\beaa}{\begin{eqnarray} }\newcommand{\eeaa}{\end{eqnarray}}
\newcommand{\bsa}{\begin{subeqnarray}}\newcommand{\esa}{\end{subeqnarray}}
\newcommand{\ba}{\begin{array}}\newcommand{\ea}{\end{array}}
\newcommand{\bit}{\begin{itemize}}\newcommand{\eit}{\end{itemize}}
\newcommand{\ben}{\begin{enumerate}}\newcommand{\een}{\end{enumerate}}
\def\lab{\label}

\def\rar{\rightarrow}

\def\al{\alpha}\def\ga{\gamma}\def\Ga{\Gamma}

\def\ka{\kappa}\def\la{\lambda}
\def\si{\sigma}\def\om{\omega}
\def\Om{\Omega}%

\def\vec#1{{\bf #1}}

\def\1{{_{1}}}\def\2{{_{2}}}

\usepackage{amsmath,amssymb,amsfonts,amsthm,epsfig}

\usepackage{epsfig}

\usepackage{graphicx}

\begin{document}

\title{On the molecular dynamics in the hurricane interactions with its environment}


\author{Gabriel Meyer\footnote{gabriel.meyer@neel.cnrs.fr - Present address}}
\address{CNRS, Institut N{\'e}el, Grenoble, 38000, France\\
and D\'epartement de Physique, Ecole Normale Sup\'erieure de Lyon, F-69342 Lyon, France}

\author{Giuseppe Vitiello\footnote{vitiello@sa.infn.it - corresponding author}}
\address{Dipartimento di Fisica ``E.R. Caianiello'' and Istituto Nazionale di Fisica Nucleare\\ Universit\'a di Salerno, I-84100 Fisciano (Salerno), Italy}

\begin{abstract}
{\it \bf Abstract}: By resorting to the Burgers model for hurricanes, we study the molecular motion involved in the hurricane dynamics. We show that the Lagrangian canonical formalism requires the inclusion of the environment degrees of freedom. This also allows the description of the motion of charged particles. In view of the role played by moist convection, cumulus and cloud water droplets in the hurricane dynamics, we discuss on the basis of symmetry considerations the role played by the molecular electrical dipoles and the formation of topologically non-trivial structures. The mechanism of energy storage and dissipation, the non-stationary time dependent Ginzburg-Landau equation and the vortex equation are studied. Finally, we discuss the fractal self-similarity properties of hurricanes.
$$  $$


\end{abstract}

%

\maketitle

\section{Introduction}

Tropical cyclones, also called hurricanes in the Atlantic ocean, are deadly large-scale storms which form over warm ocean waters in tropical regions. They have been studied for decades, and much progress in their understanding has been made through observational and theoretical study \cite{Ooyama,hurricane}. Certain factors, such as the ocean surface temperature, troposphere humidity, and Coriolis parameter, are known to be crucial for their genesis \cite{Gray}. But a consensus on the physical processes involved in tropical cyclone formation and intensification is still lacking \cite{Montgomery1992,Montgomery2014,Tory}. It is known that heat and angular momentum are two main sources of energy \cite{hurricane} for hurricanes, which in crossing the ocean may be getting stronger on the way. By reaching regions where the ocean's surface is warmer, the hurricane may indeed get more heat. In such regions, more water will evaporate from the ocean. As the vapor flows upward, it cools down and eventually condenses thereby releasing latent heat. The hurricane may get more angular momentum by coming across wind currents blowing in the right directions and sucking them in. Eliassen \cite{Eliassen} and many after him \cite{Ooyama} have used his balanced vortex model for numerical simulations. Highly non-trivial non-equilibrium thermodynamic and hydrodynamic processes are involved at a molecular level in the hurricane formation and evolution. In this paper our goal is to study phenomena of energy storage and dissipation, symmetry and topology properties at a molecular level by resorting to the Lagrangian canonical formalism of field theory. We show how some basic aspects of the underlying molecular dynamics are actually implied by specific features of the the Burgers model \cite{Burgers}. The mathematical consistency of the canonical formalism requires the inclusions of the environment degrees of freedom, which agrees with the computational strategy in hurricane studies stressing the role of the hurricane-environment interaction \cite{hurricane}. This also allows to shed some light on the dynamics of charged particle. Symmetry properties and gauge transformations in the dynamics of the molecules in moist convection, cumulus and cloud water droplets are  then considered and molecular electric dipoles in cloud droplets are studied. Criticality of the non-equilibrium dynamics and temperature dependence is analyzed by use of the non-stationary time dependent Ginzburg-Landau (GL) equation, which shows how topologically non-trivial structures (vortices) emerge out of the non-stationary fluid dynamics regime. We also show that interesting information on the fractal self-similarity properties of hurricanes may be derived.

The plan of the paper is the following. In Section 2 we shortly summarize the Burgers model and show how the canonical formalism requires the doubling of the degrees of freedom in order to include the environment in our discussion. The hurricane molecular background is considered in Section 3, where we also show that charged molecules and ions may be included in our formalism. Some noncommutative Brownian motion properties are also derived. The spontaneous breakdown of the rotational molecular dipole symmetry is considered in Section 4, where it is shown how the polarization current contributes to the Maxwell equation for the electromagnetic field. The self-focusing propagation of this field, phenomenologically well studied in beam propagation in nonlinear medium, is then discussed. Energy storage and energy dissipation at a molecular level is analyzed in the free energy minimization condition in the quasi-stationary approximation. The Ginzburg-Landau  non-stationary fluid dynamics regime and the derivation of the vortex equation is discussed in Section 5.  Temperature dependence of the vortex solution in the criticality region is also commented upon. In Section 6 we present concluding remarks and discuss the fractal self-similar properties of hurricanes emerging from the Burgers model.  Some mathematical details are presented in the Appendices A and  B.

\section{The Model}

The Burgers model \cite{Burgers} for hurricane describes how a vortex may stretch so as to concentrate its vorticity in a smaller region. In this and in the following Section we briefly summarize some of the model features and show how they can be treated within the Lagrangian canonical formalism and how they reflect on the underlying molecular dynamics.

The velocity field in cylindrical coordinates is obtained from the Navier-Stokes equation (see the Appendix A):

\begin{equation}
\left\{
\begin{array}{llc}
u_z=2Cz \\
u_r=-Cr \\
u_{\theta}=\frac{\Gamma}{2\pi r}[1-\exp(\frac{-r^2}{2\delta^2(t)})]
\end{array}
\right.
\label{velocity}
\end{equation}

\noindent The flow is a combination of two motions:
\begin{itemize}
\item[--] There is an irrotational part with radial and vertical components $u_r$ and $u_z$, which is controlled by the parameter C. In general, C can be of any sign. Considering only the region $z>0$ (place the ground at $z=0$), we can see that $C>0$ corresponds to a radially inward and vertically upward flow, whereas $C<0$ corresponds to a radially outward and vertically downward flow. As mentioned in the introduction, in the core of a hurricane, there is a vertical current of warm and wet air going upward. $C$ depends in general on the temperature $T$, the pressure, the density and viscosity of the medium, its molecular composition, the density and charge of present particles, atoms, molecules, ions, etc.; variables which    in turn depend on space ${\bf r}$ and time $t$. Thus, by $C$ we actually mean its effective value $C({\bf r} ,t)$ resulting from all of its dependence on the said variables. In the following, for definiteness we will consider the case $C>0$ and the approximation of constant $C$ over a relatively wide space region and relatively long time intervals. With due changes, our discussion and conclusions are not substantially affected by the choice $C<0$.

\item[--] The rotational part, on the other hand, is contained in the angular velocity $u_\theta$. It describes an axis-symmetric and vertically uniform vortex flow with strength and size controlled by parameters $\Gamma$ and $\delta$ respectively.
\end{itemize}

$\Gamma$ is equal to the circulation along a horizontal circle with infinite radius or, equivalently, to the flux of vorticity across the entire xy-plane :

\begin{equation}
\Gamma=\lim_{r \to \infty}  \oint_{\mathcal{C}_r}  u_{\theta} \, r\,d \theta = \iint_{xy\scriptsize\textrm{-plane}} \om \, dS
\end{equation}
where  the vorticity $\om$ is Gaussian shaped (Appendix A):
\begin{equation} \lab{omt}
\om=\frac{1}{r}\frac{d}{dr}(ru_\theta)=\frac{\Gamma}{2\pi\delta^2}\exp(\frac{-r^2}{2\delta^2})~.
\end{equation}

\begin{figure}[h!]
\centering\includegraphics[scale=0.45]{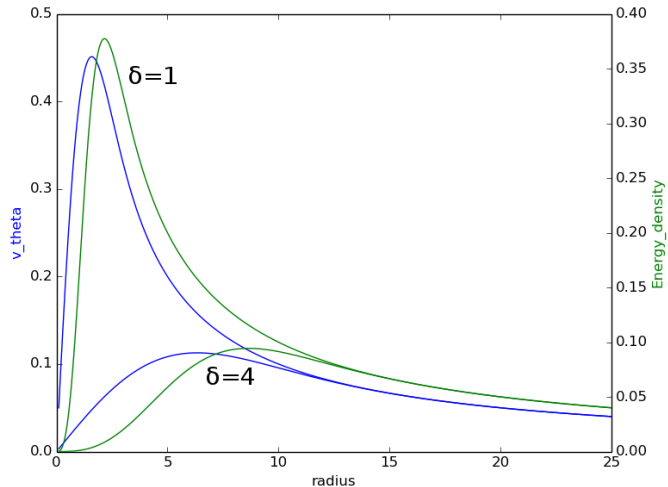}
\centering\caption{\small \noindent  $u_\theta$ and $E_{\textrm density}$ is ploted for two different values of $\delta$. The energy maximum occurs for a slightly larger value than the velocity maximum.}
\label{radius}
\end{figure}

\begin{figure}[h!]
\centering\includegraphics[scale=0.45]{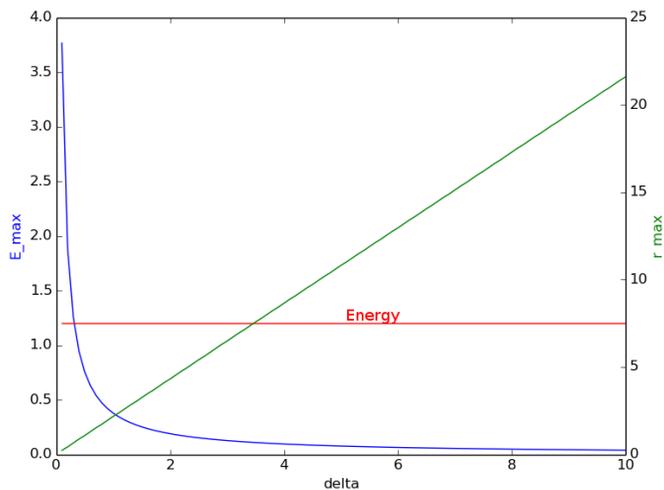}
\centering\caption{\small \noindent Plot of the energy maximum (blue) and radius where the maximum occurs (green) for different values of $\delta$. The maximum of energy increases wihout bond for small $\delta$
and the radius at the maximum increases linearly with $\delta$ (like $\approx 2\delta$). The energy integrated over five $\delta$ (red) is shown to be clearly independant of $\delta$.}
\label{delta1}
\end{figure}

The energy associated to the vortex flow (per vertical unit) is :
\begin{equation}
E_v= \int  u_\theta^2 \,\pi r \, dr = \frac{\Gamma^2}{4\pi} \int_0^\delta \frac{1}{r}[1-\exp(\frac{-r^2}{2\delta^2})]^2 \, dr
\label{energy}
\end{equation}
This integral is well defined in the limit $r\to 0$ but it diverges as $\ln(r)$ for $r\to \infty$. Therefore, the natural cutoff $\delta$ is introduced in the model. The necessity of the cutoff indicates that this model is only valid locally. We verified numerically (with python) how the energy (\ref{energy}) depends on $\delta$. In figure \ref{radius}, we show, for two values of $\delta$, the velocity and energy distribution as a function of the distance from the axis of the vortex. In figure \ref{delta1}, the maximum of energy as well as the radius at which this maximum occurs, are shown as a function of the vortex size. The integrated energy is clearly independent of $\delta$.
The energy density is instead dependent on $\delta$ as follows:
\begin{equation}
E_{\textrm density}=\frac{\Gamma^2}{4\pi r}[1-\exp(\frac{-r^2}{2\delta^2})]^2
~.
\end{equation}
The storage or dissipation of energy by hurricanes in the course of their time evolution is of course a question of great practical interest.

We focus now our attention on $\delta$ since it plays an important role in the dynamics described above.  As implied by the Navier-Stokes equation (see \cite{hurricane,Burgers} and the Appendix \ref{A}) it  evolves in time according to:
\begin{equation}
\delta^2(t)=\frac{\nu}{C}+(\delta_0^2-\frac{\nu}{C})e^{-2Ct}
\label{delta}
\end{equation}
where $\delta_0$ is the initial value of $\delta$, $\nu$ is the viscosity and  $\delta^2 (\infty) = \nu/C$ at $t = \infty$. The sign of $\delta^2(t)-{\nu}/{C}$ is determined by the sign of $\delta_0^2-{\nu}/{C}$ (in principle $\delta_0^2$ can be larger or smaller than ${\nu}/{C}$).
There is a competition between the viscosity, which tends to make the vortex diffuse, and the stretching flow, which tends to make the vortex more concentrated around its axis. Eq,~(\ref{delta}) is thus a central result in the Burgers model. Our task is now to obtain in the canonical formalism the Lagrangian from which it can be derived. The motivation is that this will provide information on the underlying molecular dynamics in the hurricane-environment interactions.

Define $\varrho^2(t)= |\delta^2(t)-{\nu}/{C}|$, $\varrho^2_0 =|\delta_0^2-{\nu}/{C}|$.
Equation (\ref{delta}) can be then written as
\bea \lab{spiral}
 \varrho(\theta) =\varrho_0\,e^{-\theta}
\eea
where the notation is $\theta \equiv \theta(t) = C\,t$. The associated parametric equations in polar coordinates 
are
\bea \lab{losp2.6} \xi &=&\varrho(\theta) \, \cos \theta =\varrho_0 \,e^{- \, \theta}\, \cos \theta ~, \\
  \eta &=&\varrho(\theta) \, \sin \theta =- \varrho_0 \,e^{- \, \theta}\, \sin \theta  \lab{losp2.6b} ~.
\eea

The completeness of the (hyperbolic) basis $\{e^{-  \theta}, \, e^{+   \theta} \}$ requires that both elements $q = e^{\pm \, \theta}$ must be taken into account in the complex $z$-plane $z =  \xi  + i\, \eta$.
We thus consider both the points  $z_1 = \varrho_0 \, e^{-  \theta} \, e^{- \,i \, \theta}$ and $z_2= \varrho_0 \, e^{+  \theta} \, e^{+ \, i \, \theta}$.  We consider for convenience both signs also for the imaginary exponent $ \pm i \, \theta$.

We observe now that $z_1 (t) = \varrho_0 \,  \, e^{- \,i \, \Om \, t}\, e^{- C t }$ and $z_2  (t) = \varrho_0 \,  \, e^{+ \,i \, \Om\,t}\, e^{ + C \, t }$  solve the equations
\bea \lab{losp2.10} m \, \ddot{z}_1 \, + \, \ga \, \dot{z}_1 \, + \, \kappa \, z_1  &=& 0 ~,\\ \lab{losp2.10b}
m \, \ddot{z}_2  \, - \, \ga \, \dot{z}_2 \, + \, \kappa \, z_2  &=& 0 ~,
\eea
respectively, with positive real $m$, $\ga$ and $\kappa$, $C \equiv {\ga}/{2 \, m}$
and $\Om^2 \equiv C^2 =  (1/m)(\ka- \ga^2 /4m)$, with $\ka > \ga^2 /4m\,$; ``dot" denotes derivative with respect to $t$. We observe that no new parameters have been introduced through Eqs.~(\ref{losp2.10}) and (\ref{losp2.10b}). In fact the ratios $\ga/m$ and $\kappa/m$ are proportional to $C$ and $C^2$, respectively.

By putting $x(t) \equiv [z_1 (t) + z^{*}_2 (-t)]/2 $ and  $y(t) \equiv [z^{*}_1 (-t) + z_2 (t)]/2$, Eqs.~(\ref{losp2.10}) and (\ref{losp2.10b}) reduce to
the couple of damped and amplified harmonic oscillators (the Bateman system \cite{Bateman,Celeghini:1992yv}):
\bea \lab{xy1}
m \ddot x + \gamma \dot x + k x  &=& 0  ~~,\\
\lab{xy2}
m \ddot y - \gamma \dot y + k y  &=& 0 ~~,
\eea
respectively. Note that $C = \gamma/2m$ controls the dissipation (amplification) terms in (\ref{xy1}) and (\ref{xy2}). Eq.~(\ref{xy2}) ((\ref{xy1})) is the time-reversed image ($\ga \rar - \ga$) of (\ref{xy1}) ((\ref{xy2})), whose physical meaning is  that total energy is conserved.  The global system $(x-y)$ is a indeed a closed system.

The Lagrangian from which Eqs.~(\ref{xy1}) and (\ref{xy2}), i.e., equivalently, Eq.~(\ref{delta}), are derived is given by \cite{Celeghini:1992yv}
\bea \lab{Lxy}
 L &=& m\dot x \dot y + {\gamma\over 2} (x \dot y -\dot x y) - k x\,y ~.
\eea
The Hamiltonian is
\bea
\lab{Hxy}
H &=&  {{1}\over{m}} p_{x} p_{y} +
{{1}\over{2m}}\gamma
\left ( y p_{y} - x p_{x} \right ) + \left ( k -
{{\gamma^{2}}\over{4 m}} \right ) x \,y ~,
\eea
with conjugate momenta $p_{x} = \partial L/\partial \dot x = m \dot y - ({\gamma}/{2}) y$, and $p_{y} = \partial L/\partial \dot y  =
m \dot x + ({\gamma}/{2}) x$.  $\Omega$  is the common frequency of the two oscillators. Note that without introducing both the modes, $x$ and $y$,  the conjugate momenta cannot be defined.  The canonical formalism can only describe closed systems. The $y$ system may be seen as playing indeed the role of the environment for the $x$ system (or vice-versa). We see thus how important it is to consider the role of the interaction with the environment, as indeed stressed in  hurricane studies \cite{hurricane}.

In conclusion, we have expressed the dynamical content of  Eq.~(\ref{delta}) of the Burgers model by the Lagangian (\ref{Lxy})  and the Hamiltonian (\ref{Hxy}).

\section{The molecular background}

In the analysis presented in the previous Section  we have considered some of the  (fluid dynamics) properties of the Burgers model of hurricanes, focusing in particular on the model features expressed by Eq.~(\ref{delta}). As mentioned in the introduction the hurricane exchanges energy by interacting at a molecular level with the air molecules (atmosphere, mostly composed of nitrogen ($N_2$) and oxygen ($O_2$), but also carbon dioxyde ($CO_2$) and Argon ($Ar$)) and water molecules largely present in the environment; moist convection, cumulus and water droplets in the clouds playing an important role in the hurricane dynamics \cite{hurricane}. The domain corresponding to the surface
$\varrho^2(t)=|\delta^2(t)-{\nu}/{C}|$ has indeed blurred borders, i.e. not sharply defined linear dimension $\varrho(t)$, due to fluctuating molecular movements arising from the hurricane interaction with the environment molecules. The aim of this Section is to study such a molecular motion in terms of the $x$ and $y$ variables introduced above.

We may represent these brownian-like molecular agitations at the border by two coordinates $x_+ (t)$ (going forward in
time) and $x_- (t)$ (going backward in time)
\bea
\lab{xplus}
x_{\pm} \equiv x \pm \frac{y}{2} \,.
\eea
Eqs.~(\ref{xy1}) and (\ref{xy2})  are then rewritten as
\bea \lab{eqplus}
m \ddot x_+ + \gamma \dot x_- + k x_+  &=& 0  ~~, \\
\lab{eqminus}
m \ddot x_- + \gamma \dot x_+ + k x_-  &=& 0  ~~.
\eea
The Lagrangian (\ref{Lxy}) and the Hamiltonian (\ref{Hxy}) become
\bea \lab{Lplus}
L = {m\over 2} \dot x_+^2 - {k\over 2} x_+^2 - {m\over 2} \dot x_-^2 + {k\over 2} x_-^2 + {\gamma\over 2} \, ({\dot x_+} x_- - {\dot x_-}x_+),
\eea
\bea
H = H_+ - H_- = {1 \over 2m} (p_+  - {\gamma\over 2}x_-)^2
+ {k\over 2} x_+^2 - {1 \over 2m} (p_- + {\gamma\over 2}x_+)^2 -
{k\over 2} x_-^2. \lab{Hplmin}
\eea
respectively, with conjugate momenta $p_+ = m \dot x_+ + ({\gamma}/{2}) x_- ~,~~   p_- = -m\dot x_- - ({\gamma}/{2}) x_+$.
It is interesting that the damping manifests itself as a correction in
the kinetic energy and that the dissipative constant acts as a coupling
between $x_+$ and $x_-$ (cf. Eq.~(\ref{Lplus})).

From Eq.~(\ref{Hplmin}) we obtain the forward and backward in time velocities $v_\pm$ 
\be v_{\pm }={\partial {\cal H} \over
\partial p_{\pm }} =\pm \, \frac{1}{m}( p_\pm \mp \frac{\ga}{2}\,
x_\mp) ~. \label{(9)} \ee
By using as usual $p_\pm = -i\hbar d/d x_\pm$, 
we have
\be [v_+,v_-]=i\hbar \,{\ga\over m^2}~.  \label{(10)} \ee
Provided that we set $ \gamma \equiv q\,B_3/ c$, Eq.~(\ref{(10)}) is similar to the usual
commutation relations for the velocities ${\bf v}=({\bf
p}- q{\bf A}/c)/m$ of a charged particle moving in a magnetic field
${\bf B}$; i.e., $[v_1,v_2]=(i\hbar \, q\,B_3/m^2 c)$ \cite{Landau}.  It is also remarkable that  the (Brownian motion) friction coefficient $\ga$ in  (\ref{(10)}) induces a phase interference between forward and backward
motion \cite{Sivasubramanian:2003xy} analogous to the Bohm-Aharonov phase interference for charged particles. The commutator (\ref{(10)}) also implies a topologically non-trivial (noncommutative) geometry in the  $(v_+, v_-)$ plane, reflecting of course the non-trivial topology of the hurricane vortex.

These remarks and the fact that a non-vanishing density of ionized particles may be present since the hurricane vorticity and other forces may cause molecule ionization,  suggest to us to write  Eq.~(\ref{Hplmin})  as
\bea \lab{HA}
H = {H_+} - {H_-} = {1 \over 2m} (p_+ - {q_+ \over{c}}{A_+})^2  +
{q_+}{\Phi}_+
- {1 \over 2m} (p_- + {q_- \over{c}} A_-)^2 + {q_-}{\Phi}_- ,
\eea
where the notation is
\bea
A_+ \equiv  {B\over 2}  x_- ~, \qquad  A_- \equiv  {B\over 2}  x_+ ~, ~~{\rm with} ~~~B \equiv
{c\over{q}}\gamma ,
\lab{2.21}
\eea
namely, $(\gamma/2) x_{\pm}$ in Eq.~(\ref{Hplmin}) are represented as the components $A_+ \equiv A_1$ and $A_- \equiv A_2$ of the electromagnetic (em) vector potential ${\bf A}$ \cite{Sivasubramanian:2003xy,Chern} with the magnetic field ${\bf  B} = {\bf \nabla} \times {\bf A} = - B {\bf {\hat 3}}$. The Hamiltonian  $H$ then describes two particles with opposite charge $q_+ = - q_- = q$
in the (oscillator) potential
$\Phi \equiv {(k/ 2q)}({x_+}^2 - {x_-}^2) \equiv {\Phi}_+ - {\Phi}_- ~,
{\Phi}_\pm \equiv {(k/ 2q)}{x_\pm}^{2}$,
and in the constant magnetic field $\bf B$. Eqs.~(\ref{HA}) and (\ref{2.21}) show that one particle moves in the em field whose source is the other particle.
Eqs.(\ref{eqplus}) and (\ref{eqminus}) are recognized to describe nothing but the Lorentz forces on
particles with charge $q_+ = - q_- = q$, in the magnetic field ${\bf B} = - B  {\bf {\hat 3}}$ and in the electric field
${\bf E} = - {\bf \nabla}  \Phi $:
\bea
{\cal F}_{i} = m\ddot x_i = {q_i}[E_i + {1\over{c}}({\bf v} \times
{\bf B})_i] ~, \lab{Ltz}
\eea
with the notation $i = 1,2 \equiv +, - $, respectively, and   ${\bf v} = (\dot x_+ , \dot x_- , 0)$.

The conclusion is that the Hamiltonian (\ref{Hplmin})  accounts for the dynamics of molecular motion and, with the identification (\ref{2.21}), also of  charged particles motion in the hurricane-environment interaction.
In the following Section we continue the analysis of the molecular dynamics by considering the role of the breakdown of the rotational symmetry of molecular electrical dipoles.

\section{The spontaneous breakdown of symmetry}

It is well known that a weak perturbation acting on a system may trigger the breakdown of the symmetry of the dynamics of the system's elementary components \cite{Anderson:1984a,ITZ,Umezawa:1993yq,DifettiBook}. It is well established that as a consequence of the symmetry breakdown, the system components undergo an {\it in phase}, or coherent collective motion resulting in ordered states. The ordering of the elementary  components  is generated by long range correlations among them.
These correlations arise spontaneously as a dynamical effect of the process of the symmetry breakdown (the Goldstone theorem)  \cite{Umezawa:1993yq,DifettiBook}.
The degree of ordering is described by a quantity, called order parameter, characterizing the macroscopic behavior of the system. The change of scale, from the dynamics of the elementary components to the system's macroscopic behavior, is thus obtained.
It can be shown \cite{Matsumoto:1975fi,Matsumoto:1975rp} that the operatorial formalism leads to classical field equations and observables. In our discussion in the following we will thus always refer to such a classical level of description, which emerges out of the elementary (molecular) component dynamics.

The phenomenon of the spontaneous  breakdown of symmetry (SBS) is so widely diffused in natural phenomena that it acquires a paradigmatic character \cite{Umezawa:1993yq,DifettiBook}.
In the present paper, our standpoint is that in the region where the hurricane is generated,  a weak  perturbation, e.g. the em field generated by ionized particles, or perturbations of other kind (mechanical, chemical, etc.), may break the rotational symmetry of the electric dipole of the water molecules present in that region. The role played by cloud water droplets, cumulus and moist convection in hurricane genesis, formation and evolution has been indeed much studied in the literature with numerical and analytical analysis (see e.g. \cite{Ooyama,hurricane}). We thus consider the non-homogeneous polarization density ${\cal P} ({\bf r}, t)$ that may arise as a dynamical effect of the SBS. The consequent collective correlations among the in phase oscillating molecular dipoles are described by dipole fields (the Nambu-Goldstone (NG) fields) whose coherent condensation in the ground state manifests itself in the classical, macroscopic properties and behavior of the system.
Let $\rho ({\bf r}, t)$ denote the charge density and let $\delta$ be the (average) dipole length. Then we obtain \cite{DelGiudice:1986} $2{\cal P} ({\bf r}, t) = \rho ({\bf r}, t) \, \delta$.

We observe that since it is impossible to change by a
constant amount  the phase of the dipole field simultaneously at every space-point, the symmetry under rotations by a constant phase $\la$ around the dipole axis (the global U(1) symmetry) is broken as well.
One can show that under quite general
conditions the order parameter $v ({\bf r}, t)$ is proportional to ${\cal P} ({\bf r}, t)$: $|v ({\bf r}, t)|^2 \propto 2{\cal P} ({\bf r}, t) = \rho ({\bf r}, t) \, \delta$ \cite{DelGiudice:1986}, and by adopting a standard recipe \cite{Umezawa:1993yq,DifettiBook,DelGiudice:1986}
we describe the order parameter in terms of the charge density
wave function $\sigma ({\bf r}, t)$:
\be\lab{sigma}
\sigma ({\bf r}, t) = \sqrt{\rho ({\bf r}, t)}\, e^{i\varphi({\bf r}, t)}  ~,
\ee
with real $\varphi({\bf r}, t)$. In the U(1) SBS, the $\varphi({\bf r}, t)$ field plays the role of the NG wave field and the transformation $\varphi ({\bf r}, t) \rightarrow \varphi ({\bf r}, t) + ({q}/{\hbar \,c})\, \la  ({\bf r}, t)$, inducing the phase transformation for $\sigma ({\bf r}, t)$, describes its coherent boson condensation process \cite{Umezawa:1993yq,DelGiudice:1985}.
In full generality, one may consider the transformation
\be\lab{2.20bt}
\varphi ({\bf r}, t) \rightarrow \varphi ({\bf r}, t) +  f({\bf r}, t) ,
\ee
with $f({\bf r}, t)$ playing the role
of a form factor in the $\varphi$ non-homogeneous condensation. Observable effects and topologically non-trivial structures are obtained when $f  ({\bf r}, t)$  carries  topological singularities \cite{Umezawa:1993yq,DifettiBook,Matsumoto:1975fi,Matsumoto:1975rp}.
For example, vortices  appear by using
\be \lab{atg} f(x) = \arctan \left(\frac{x_{2}}{x_{1}}\right)~,
\ee
which indeed  carries a singularity on the line
$r = 0$, with $r^{2} = x^{2}_{1} + x^{2}_{2}$.

Consider now the em field gauge transformation ${\rm\bf A} ({\bf r}, t) \rightarrow {\rm\bf A}' = {\rm\bf A} ({\bf r}, t) + {\mbox{\boldmath $\nabla$}} \la ({\bf r}, t)$.
When using $\varphi$ as gauge function in ${\rm\bf A} \rightarrow {\rm\bf A}' = {\rm\bf A} + (\hbar\, c /q) {\mbox{\boldmath $\nabla$}} \varphi$,  the  transformation ${\rm\bf A}' \rightarrow {\rm\bf A}' + {\mbox{\boldmath $\nabla$}}\la$  is induced by $\varphi ({\bf r}, t) \rightarrow \varphi ({\bf r}, t) + ({q}/{\hbar \,c})\, \la  ({\bf r}, t)$. 
We adopt the gauge condition ${\mbox{\boldmath $\nabla$}}\cdot {\rm\bf A} = 0$ (${\mbox{\boldmath $\nabla$}}\cdot {\rm\bf A}' = 0$), which requires
${\mbox{\boldmath $\nabla$}^2} \varphi ({\bf r}, t) = 0$ and ${\mbox{\boldmath $\nabla$}^2} \la  ({\bf r}, t) = 0$ (and also ${\mbox{\boldmath $\nabla$}^2} f({\bf r}, t) = 0$).

One can show that the Maxwell equation $\Box {\bf A} ({\bf r},t) = {\bf J}_p ({\bf r},t)$, where the polarization current ${\bf J}_p = \rho ({\bf r}, t) {\bf v}$, with ${\bf v} = \partial$${\mbox{\boldmath $\delta$}}$/$\partial t$, is assumed to be the only relevant current, may be rewritten as
\bea \lab{em1}
( \Box + M^2) {\bf A} (x) = \frac{2 \, {\cal P}}{m\, \delta} \,\hbar \, {\mbox{\boldmath $\nabla$}} \varphi ({\bf r}, t) .
\eea
where $M^2 \equiv {2 q \, {\cal P}}/(m \, \delta\,c)$. In obtaining (\ref{em1}) we used
\bea \lab{J}
{\bf J}_p ({\bf r}, t) = \frac{1}{m} \rho ({\bf r}, t)(\hbar \, {\mbox{\boldmath $\nabla$}} \varphi ({\bf r}, t) - \frac{q}{c}\, {\rm\bf A}({\bf r}, t) ),
\eea
derived in a standard fashion by using (\ref{sigma}) in the definition of ${\bf J}_p$ and the usual minimal coupling derivative $(-i\hbar\mbox{\boldmath$\nabla$}-q {\rm\bf A})$. From (\ref{J}) we see that
${m}\, {\rm\bf v} =   \hbar \, {\mbox{\boldmath $\nabla$}} \varphi - ({q}/{c})\, \rm\bf A$, with ${\mbox{\boldmath $\nabla$}} \cdot {\rm\bf v} = 0$ and ${\mbox{\boldmath $\nabla$}} \cdot {\bf J}_p = 0$. In condensed matter physics,  $(q/ mc)\,\rho ({\bf r}, t)\rm\bf A({\bf r}, t))$  is the Meissner current term and $({1}/{m}) \rho ({\bf r}, t){\hbar \,\mbox{\boldmath $\nabla$}} \varphi({\bf r}, t)$ is the boson current.

The SBS mechanism thus dynamically produces the non-vanishing mass $M$ for the em gauge field $\bf A$ (the well known Anderson-Higgs-Kibble mechanism) \cite{Anderson:1984a,ITZ,Umezawa:1993yq,DifettiBook,hig},  implying its damping  by a factor  $\propto \exp(- c\,M\, d/ \hbar)$ and its self-focusing propagation with a transversal size of the order of $d \propto 1/M$ \cite{Matsumoto:1975fi,Matsumoto:1975rp,DelGiudice:1986}. Self-focusing propagation of em fields is a well studied phenomenon in classical nonlinear optics \cite{Marburgher,Chiao,Shen} and it has been shown \cite{DelGiudice:1986} that the present  analysis accounts for phenomenological aspects of em wave propagation in nonlinear non-homogeneous media, as it is indeed the medium of the hurricane highly nonlinear turbulent fluid dynamics.

We observe that the dipole of the water molecule is due to the displacement of a pair of electrons; thus we can use $q = 2e$ and $m = 2m_e$ in the expression for $M^2$ obtained above. The value of the polarization density ${\cal P}$ is then $2\,e \,\delta/c$ times the number $n$ of oriented dipoles, which in the extreme case where all dipoles are oriented is given by $n = N/18$, where $N$ is the Avogadro number.

In a unit volume, by restoring the Coulomb constant $1/ 4 \pi \epsilon_0 $, we then obtain $M\,c^2 = \hbar \,c\,(\,16 \, \pi \, r_{0}\, n)^{1/2} \approx 13.60~~ {\rm eV}$,
where  $r_{0}= (1/4 \,\pi \, \epsilon_0)(e^{2}/ m_e \, c^2) = 2,8 \times 10^{-15}$ m is the ``classical radius" of the electron, and we used $\hbar =0,6582 \times 10^{-15}$ eV sec;  $c$ is the light velocity and the unit volume is $1$ m$^3$. We see that the mass acquired by the self-focusing em field in
such a case corresponds to the energy approximately equal to the hydrogen ionization energy. A single photon with energy $\hbar\,\nu < M\,c^2 $ is thus not able to propagate through the correlated medium. A highly energetic photon ($\hbar\,\nu \gg M\,c^2 $) would produce instead the breakdown of the correlation modes, thus restoring the massless em field propagation (symmetry restoration).

In general the polarization density values are smaller than the maximum value above considered. There is, however, an interesting mechanism of energy storage in coherent states. Energies from different processes (mechanical, chemical, electromagnetic), less than the hydrogen ionization energy, could be stored indeed as polarization modes in the coherent state, provided that they can produce excitations in the correlated medium. When the accumulated energy is enough to reach the ionization value, the whole polarization mode would manifest as an em field propagating as in (\ref{em1}) with the corresponding energy value
$M\,c^2$.
It is indeed possible to show that the condensation of the polarization modes depends on the energy exchanges between the system and the environment (energy supply and energy dissipation), eventually resulting into entropy variations. In fact, in the quasi-stationary approximation, the  minimization of the free energy $ {\cal F}$,   $ d {\cal F} = d E - {1\over{\beta}} d S=0 $, with $\beta = 1/k_B T$, implies that the rate of change of the number $N_{\kappa}$ of condensate polarization modes is  related to the changes of the `internal energy' $d E$ and entropy $d S$ by the relation \cite{Celeghini:1992yv,Umezawa:1993yq,DifettiBook}
\be \lab{var} d E= \sum_{\kappa} E_{\kappa} \,
\dot{N}_{\kappa} (t)d t = {1\over{\beta}} d S ~, \ee
with  ${\dot N}_{\kappa} \equiv dN_{\kappa}/dt$. As usual, the exchanged heat is $dQ = ({1/{\beta}}) dS$.  Eq.~(\ref{var}) expresses the first
principle of thermodynamics for a system coupled with environment
at temperature $T$ and in the absence of mechanical work. It shows that, in quasi-stationary conditions, the balance $ d {\cal F} = 0$ is preserved by compensation of  changes in the entropy with changes in the system internal energy (and viceversa). The rate of change $dN_{\kappa}/dt$  of condensate polarization modes may turn into internal energy storage or dissipation in the form of heat $dQ$, or, in the presence of em field and topological non-trivial geometry (topological singularity at $r =0$), as massive photon energy.

For our study of the molecular dynamics involved in the hurricane formation and evolution,  most interesting is, however, the non-equilibrium dynamics characterized by criticality and phase transitions (assumed anyway to occur between thermodynamic equilibrium states to which the
fluctuation theorem \cite{Crooks} applies). Such a non-equilibrium dynamics can be analyzed  by use of the non-stationary time dependent Ginzburg-Landau (GL) equation \cite{Obinata}, which describes the rate at which the system approaches the stationary regime at the minimum of the free energy.
It is indeed in such a non-stationary fluid dynamics regime that the topologically non-trivial structures (vortices) considered above emerge \cite{Obinata}.
We discuss the non-stationary time dependent Ginzburg-Landau equation in the following Section.

\section{The Ginzburg-Landau non-stationary regime}

We are interested in the variations of the wave function density   $\sigma ({\bf r}, t)$ occurring in the non-equilibrium
phase transition processes.

Denote by $F (\sigma, \sigma^{*}, \rm\bf A)$ the free energy density Ginzburg-Landau (GL) functional, whose explicit form depends on the particular  model one adopts.  In general it contains  a kinetic energy term $(-i\hbar\mbox{\boldmath$\nabla$}- (q/c) {\rm\bf A})^2/2m$, some potential term and it is a non-linear functional of the fields.
Extremizing $F (\sigma, \sigma^{*}, \rm\bf A)$ with respect to $\si^{*} ({\bf r}, t)$ gives the stationary Ginzburg-Landau (GL) equation: ${\partial F }/{\partial\sigma^*} = 0$.
In full generality, we may write:
\begin{eqnarray} \lab{SGL}
	\frac{\partial F }{\partial\sigma^*}
\, \equiv \,
	\left[\frac{1}{2 m }(-i\hbar\mbox{\boldmath$\nabla$}-  {q\over c } {\rm\bf A})^2
	+\mu^2+\la|\sigma ({\bf r}, t)|^2\right]\sigma ({\bf r}, t),
\end{eqnarray}
with $\mu^2(T)$ and $\la(T)$ acting as a mass term and a positive coupling constant term, respectively.
Their explicit expressions depend on the system under study.
In the r.h.s. of Eq.~(\ref{SGL}), the quantity $(\mu^2+\la|\sigma|^2)\sigma$  may be considered as derived from the potential $V(\sigma, \sigma^*) = - (\la/2)(|\sigma|^2 + \mu^2/\la )^2$. The value of $\sigma$ minimizing the potential $V(\sigma, \sigma^*)$ is zero (disordered or symmetric ground state) for $\mu^2 \ge 0$, and non-zero for $\mu^2 < 0$, with $|\sigma|^2 = - \mu^2/\la \neq 0$ (the ordered or asymmetric ground state, with $\sigma$ playing the role of the order parameter of the breakdown of the phase symmetry). The dependence on a critical temperature $T_C$ may be taken into account through the dependence of $\mu^2$ on temperature in a form, e.g., proportional to $(T - T_C)/T$, assuming T varies linearly with time near the transition region. In the Haken interpretation and language \cite{Haken:1984a}, by tuning $\mu^2$, the pump parameter,
the system may be carried in the lasering region; in the lasering process (the phase transition)  between
the disordered and the ordered (laser) state,
$|\sigma|^2$ changes in time going from zero to a non-vanishing value proportional to
$\mu^2$.

In the non-stationary regime, ${\partial F }/{\partial\sigma^*}$ is non-vanishing and expresses the rate at which $\sigma ({\bf r}, t)$ approaches its stationary value at the minimum of the free energy. The generalized time dependent GL  (TDGL) equation is:
\begin{eqnarray}\lab{SGL12}
	{i\hbar} \frac{\partial \sigma}{\partial t}
= \hat{H}\sigma -\frac{i}{\gamma}\frac{\partial F }{\partial \sigma^{*}} ~.
\end{eqnarray}
The second term in the r.h.s. of Eq.~(\ref{SGL12})  describes the dissipative contribution coming from incoherent relaxation processes \cite{Barybin}, with $\ga$ the relaxation parameter. In the Appendix \ref{B} some details of the relaxation process are considered. In the limit of the stationary regime  $(1/\gamma)\partial F /\partial \sigma^{*} \rightarrow 0$,  the equilibrium wave function is denoted by $\sigma^{\circ}(t) = |\sigma^{\circ}| \exp(- i\epsilon t/\hbar)$, with $\hat{H}\sigma^{\circ} = \epsilon \sigma^{\circ}$.  By using $\sigma=\sqrt{n}\exp({i\varphi})$, one can show (see the Appendix \ref{B}) \cite{Obinata} that the TDGL equation becomes
\begin{eqnarray}\lab{nssv}
\frac{1}{D_{\rm GL}}\frac{d\chi}{dt} =
	\mbox{\boldmath$\nabla$}^2\chi + \frac{1}{\xi^2_{\rm GL}}\left[(1-\chi^2)-\xi^2_{\rm GL}\left(\frac{m v }{\hbar}\right)^2\right] \chi~,
\end{eqnarray}
where $\chi=|\sigma|/|\sigma^{\circ}|\equiv(n /n^{\circ} )^{1/2}$ is the normalized wave function,  $D_{GL} \equiv \xi^2_{GL}/\tau_{GL}$, is the diffusion coefficient, with $\xi_{GL}$ and $\tau_{GL}$ denoting the GL correlation length and the GL relaxation life-time, respectively (see the Appendix \ref{B}).
Eq.~(\ref{nssv}),  for $({1}/{D_{\rm GL}}){d\chi}/{dt} = \hbar \ga \, {d\chi}/{dt} \approx 0$  \cite{DifettiBook,Manka:1990} is indeed the vortex equation.
It is associated with the singularity at $r = 0$ discussed in the previous Section. For numerical analysis and simulations of Eq.~(\ref{nssv}) see, e.g. \cite{Bettencourt} and refs. therein quoted.
One may derive from the TDGL equation (see also \cite{Manka:1990})  that,  the vortex core size is, in the first approximation, of the order $\xi^2_{\rm GL} \propto |\mu^2|^{-1}$. The vortex core thus increases as temperature $T$ increases, approaching to  $T_{C}$ from below; the vortex envelope disappears at $T_{C}$. Symmetry restoration is obtained for temperature rising above $T_{C}$.

Vice-versa, going back to $T_{C}$ from above the unstructured (fully symmetric) background activity exhibits an undefined phase at $T_{C}$, i.e. the singularity at the center line $r = 0$. The critical regime starts as $T$ goes below $T_{C}$, vortices appear, and their core shrinks as temperature decreases.

Summarizing, the envelope $\sigma ({\bf r}, t)$ is temperature dependent, $\sigma ({\bf r}, t) = \sigma ({\bf r}, t, \beta)$,  and thus it changes with $T$ \cite{Manka:1990,MankaNuclPhys}. The vortex formation starts form the singularity,  at the  vortex core, in the process of phase transition due to SBS (``where the vortex comes from'') \cite{DifettiBook,Manka:1990}.
The cooling itself is a manifestation of the process of symmetry breakdown. The non-homogeneous condensation and ordered localized patterns are lower in energy and separated by an energy gap from the symmetric state. The
phase transition is non-instantaneous and begins with cooling \cite{Manka:1990}.

In the TDGL formalism it is assumed that  a time-dependent temperature might still be defined in non-instantaneous phase transition processes. In the general picture the transition is assumed to start at the critical temperature $T_C$. Time evolution leads then to the stable configuration at the so-called ``Ginzburg temperature" $T_G$, with  $T_G < T_C$. The critical regime \cite{Bettencourt} is defined to be the one evolving in time between $T_C$ and $T_G$, and is supposed to be always a transition between different
thermodynamic equilibrium states,  regardless of
the non-equilibrium processes occurring in the course of the transition process.
Thus the general thermodynamic description derived from the
fluctuation theorem \cite{Crooks} applies to the `critical dynamics' of our system. As already mentioned,
in the process of reaching new equilibrium configurations below
the critical temperature $T_C$,
local exchanges of heat with environments turn into entropy changes of the vortex, with consequent
rearrangements of its configurations and (internal) energy density contents.

Here we remark that the departure from the stationary regime (at $T_C$),
namely the start of the critical regime, is driven by fluctuations
which can trigger a phase transition, if the necessary transition energy is provided by some external input. These ground state fluctuations turn into temperature fluctuations since in our dissipative model the ground state is in fact a thermal state \cite{DifettiBook,Manka:1990}. At the end of the critical regime (at $T_G$) the system arrives at a ``new" ground state configuration and the phase transition is thus completed. As a matter of fact, the system undergoes a continuous sequence of phase transitions going through a path, or trajectory, through the (infinitely) many coherent ground states (in each of them free energy is minimized).

Remarkably, these trajectories in the manifold of the coherent states can be shown to be classical chaotic trajectories \cite{Vitiello:2004}, showing the sensibility of the molecular dynamics and of the hurricane evolution in general to local slight changes in the environmental conditions and in the same molecular system.

\section{Concluding remarks}

In this paper we have considered critical transition processes in the molecular dynamics associated to the hurricane-environment interactions. By resorting to the Burgers hurricane modeling \cite{Burgers}, we have shown 
that by including the environment degrees of freedom, several aspects of molecular motion can be derived in the canonical Lagrangian formalism.

In completely  different contexts, e.g. condensed matter physics,  cosmology, etc., theoretical and experimental research has shown that extended objects with non trivial topology (also called ``topological defects'') appear during critical transition processes (see e.g. \cite{kib2,zurek1,Alfinito:2002a,Alfinito:2002b}) and persist for varying time intervals thereafter. Such an occurrence has been considered in our analysis in this paper. We have discussed the spontaneous breakdown of the rotational symmetry of the electrical dipoles of water molecules in the hurricane interaction with moist convection flows, cumulus and cloud water droplets.   The non-stationary states have been studied by use of the non-stationary time dependent Ginzburg-Landau equation, obtaining the vortex equation. Temperature effects and the non-equilibrium dynamics characterized by criticality and phase transitions have been discussed. We have also discussed the energy storage and energy dissipation in coherent states in connection with  heat, internal energy and entropy changes  in the molecular dynamics, so as to minimize the free energy in stationary states.

We observe that in our discussion we have always referred to the classical level of description, which is a typical feature of the spontaneous breakdown of symmetry formalism allowing the change of scale, from the microscopic level of the elementary components to the macroscopic behavior of the system.

\begin{figure}[h!]
\centering \resizebox{8cm}{!}{\includegraphics{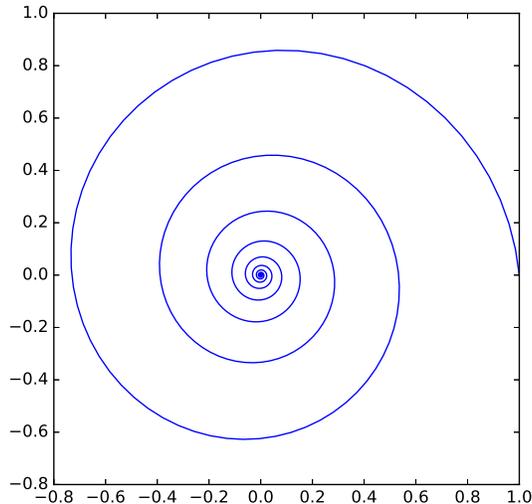}}
\caption{\small \noindent Eq.~(\ref{spiral}): the clockwise logarithmic spiral.}
\label{fig2}
\end{figure}

Finally, we remark that  hurricanes have fractal self-similarity properties. In the Burgers model,  Eq.~(\ref{spiral})  may be indeed regarded as the logarithmic spiral equation in polar coordinates (see Fig. 3) \cite{Peitgen}. In order to see this, we consider a different parametrization which does not change the results of the discussion in the previous Sections. We thus introduce one additional parameter, accounting for example of possible changes in   $\Om$, i.e.
$ \xi =\varrho_0 \,e^{- \, \theta}\, \cos \al ,    ~\eta =- \varrho_0 \,e^{- \, \theta}\, \sin \al $ (cf.  Eqs.~(\ref{losp2.6}) and (\ref{losp2.6b})). Since it is always possible to write $\al = \theta / d$, with real $d$, the only change that we get is $\Om = C / d$ and  Eq.~(\ref{spiral}) may be also rewritten as $\varrho(\al) =\varrho_0\,e^{- d\al}$.
This can be  represented in a log-log plot with abscissa $\al = \ln e^{\al}$ by the straight line of slope $d$:
\be \lab{slope}   \ln \frac{\varrho}{\varrho_0} = - d \, \al~.
\ee
The constancy of the angular coefficient $\tan^{-1} d$ represents the self-similarity property of $\varrho$ \cite{Peitgen,PLA}. The parameter $d$ is called the fractal or self-similarity dimension. Rescaling  $t \rar n \, t$ changes $\varrho/\varrho_0 = \,e^{- d\al}$ by the power $(\varrho/\varrho_0)^n$.
At time $\tau = 2\, \pi \, d/ C$ it is $(C/d)\,\tau = 2 \, \pi$ and at $t = n \,\tau$, $z_1 = \varrho_0 \, (e^{ -\,2 \, \pi d})^n $, $z_2 = \varrho_0 \, (e^{ 2 \, \pi d})^n $, with integer $n = 1,\, 2,\, 3$... Such a macroscopic self-similarity feature of the logarithmic-spiral-like hurricane, finds its correspondence in the coherent states structure at the molecular level discussed in the previous Sections. It is indeed known \cite{FractalQI,PLA,Systems} that an isomorphism exists between
self-similarity structures and $d$-deformed coherent states.

\section*{Acknowledgements}
G.M. acknowledges the Ecole Normale Sup\'erieure de Lyon for supporting in the fall 2017 his visiting the Department of Physics ``E.R.Caianiello'', University of Salerno. G.V. acknowledges partial financial support by INFN and Miur.

\section*{Declaration} Declarations of interest: none.

\appendix


\section{On the Navier-Stokes equation and vorticity }\label{A}

\noindent The Navier-Stokes equation for a fluid with fixed density $\rho$ and viscosity $\nu$ is:
\begin{equation}
\frac{\partial\vec{u}}{\partial t}+(\vec{u} \cdot \nabla) \vec{u} = \frac{1}{\rho} \nabla p + \nu \Delta \vec{u} ,
\label{NS}
\end{equation}
where $\vec{u}$ denotes the velocity field and $p$ is the pressure.
Using the vorticity $\vec{\omega}=\nabla \times \vec{u}$, the second term in the l.h.s. becomes:
\begin{equation}
(\vec{u} \cdot \nabla) \vec{u} = \nabla (\frac{1}{2}\vec{u^2})-  \vec{u}\times\vec{\omega}.
\end{equation}
Then, the rotor of (\ref{NS}) gives:
\begin{equation}
\frac{\partial\vec{\omega}}{\partial t} - \nabla\times(\vec{u}\times\vec{\omega}) = \nu \Delta \vec{\omega}.
\end{equation}
Since the fluid is incompressible, the continuity equation reduces to $\nabla \cdot \vec{u}=0$ and we have:
\begin{equation}
\nabla\times(\vec{u}\times\vec{\omega}) = (\vec{\omega}\cdot\nabla)\vec{u} - (\vec{u}\cdot\nabla)\vec{\omega}.
\end{equation}
Therefore, we obtain the following vorticity equation:
\begin{equation}
\frac{\partial\vec{\omega}}{\partial t} +  (\vec{u}\cdot\nabla)\vec{\omega} = (\vec{\omega}\cdot\nabla)\vec{u} + \nu \Delta \vec{\omega}.
\label{vorticity}
\end{equation}

Now, since we are interested in describing a vortex, it is natural to place ourselves in cylindrical coordinates, with the z-axis being the axis of the vortex. Let us look for a flow which is such that the vertical component of the velocity is equal to $u_z=2Cz$ (with C a constant), and the other components $u_r$ and $u_\theta$ are functions of $r$ only. Then, from the continuity equation, we obtain that the radial velocity must be equal to $u_r=-Cr+ {A}/{r}$, where we will choose $A=0$ for convenience. So, we have (cf. Eq.~(\ref{velocity})):
\begin{equation}
\left\{
\begin{array}{llc}
u_z=2Cz \\
u_r=-Cr
\end{array}
\right.
\label{velocity2}
\end{equation}
The fact that the velocity increases without bond with the distance from the origin indicates that this model is only valid locally.

We now use equation (\ref{vorticity}) to find the vorticity, from which we will derive the last component of the velocity field $u_\theta$. From the conditions we have imposed on the flow, it follows that the vorticity is parallel to the z-axis: $\vec{\omega}(\vec{r},t)=\omega(r,t)u_z$. Using (\ref{NS}), the vorticity equation becomes:
\begin{equation}
\frac{\partial\omega}{\partial t}=2C\omega + Cr\frac{\partial\omega}{\partial r} + \frac{\nu}{r}\frac{\partial}{\partial r}(r\frac{\partial\omega}{\partial r}).
\label{vorticity2}
\end{equation}
In the absence of incoming or outgoing flux, $C=0$ and equation (\ref{vorticity2}) reduces to the usual diffusion equation:
\begin{equation}
\frac{\partial\omega}{\partial t}= \nu\Delta\omega,
\end{equation}
which has the solution:
\begin{equation}
\omega(r,t)=\frac{\Gamma}{4\pi\nu t}\exp(\frac{-r^2}{4\nu t}).
\end{equation}
Therefore, in order to solve the case $C\neq0$, we look for a solution of the form (cf. Eq.~(\ref{omt})):
\begin{equation}
\omega(r,t)=\frac{\Gamma}{2\pi\delta^2(t)}\exp(\frac{-r^2}{2\delta^2(t)}).
\label{vorticity3}
\end{equation}
Plugging (\ref{vorticity3}) into (\ref{vorticity2}) yields:
\begin{equation}
(\delta(t)\dot{\delta}(t)+C\delta^2(t)-\nu)(2-\frac{r^2}{\delta^2(t)})=0.
\label{deltaD}
\end{equation}
Since we are looking for a solution $\delta(t)$ which is independent of $r$, we can simplify equation (\ref{deltaD}) by $2- {r^2}/{\delta^2(t)}$. Recognizing a linear differential equation of first order for $\delta^2(t)$, we get Eq.~(\ref{delta}):
\begin{equation}
\delta^2(t)=\frac{\nu}{C}+(\delta_0^2-\frac{\nu}{C})e^{-2Ct}.
\label{delta2}
\end{equation}
Finally, using:
\begin{equation}
\omega(r,t)=\frac{1}{r}\frac{\partial}{\partial r}[ru_\theta(r,t)].
\end{equation}
We obtain (cf. Eq.~(\ref{velocity})):
\begin{equation}
u_{\theta}=\frac{\Gamma}{2\pi r}[1-\exp(\frac{-r^2}{2\delta^2(t)})].
\end{equation}

\section{Time dependent Ginzburg-Landau equation}\label{B}

The continuity equation:
\begin{eqnarray}\lab{continuity}
	\frac{\partial n }{\partial t}
	+\mbox{\boldmath$\nabla$}\cdot(n {\rm\bf v} )
	=-\frac{2G}{\tau_{\rm GL}}n
\end{eqnarray}
and the relation for the chemical potentials $\mu_0$, $\mu_1$ and $\mu_2$
\begin{eqnarray}\lab{chempot}
	\mu_{0}+\mu_{1}+\mu_{2}+q \varphi+\frac{m v^2 }{2}
	+\hbar\frac{\partial \theta}{\partial t}
	=0,
\end{eqnarray}
are derived from the imaginary and real part, respectively, of Eq.~(\ref{SGL12}) by using $\sigma=\sqrt{n}\exp({i\varphi})$ and ${\bf J} \equiv q n {\rm\bf v}$. In Eq.~(\ref{chempot}) the dissipative quantity
\begin{eqnarray}\lab{chempot2}
	\mu_{2}=-\frac{\hbar}{2 \gamma n }
	\mbox{\boldmath$\nabla$}\cdot(n {\rm\bf v} )
	\equiv -\frac{\xi^2_{\rm GL}}{\tau_{\rm GL}}\frac{m}{q}
	\frac{\mbox{\boldmath$\nabla$}\cdot {\rm\bf J} }{n }~,
\end{eqnarray}
related to the relaxation parameter $\ga$, is  proportional to non-homogeneities of the condensate.
${\tau_{\rm GL}}$ and $\xi^2_{\rm GL}$ are the GL relaxation time and coherence length, respectively.
$G$ is given by
\begin{eqnarray}\lab{G}
	G=\xi^2_{\rm GL}\frac{2m}{\hbar^2}\left(\frac{m v^2 }{2}+\mu_{1}\right)
	+\frac{n }{n^{\circ} }-1~.
\end{eqnarray}
By using Eq.~(\ref{G}) and $\mbox{\boldmath$\nabla$}\cdot {\rm\bf v} = 0$, the continuity equation  (\ref{continuity}) is  rewritten as
\begin{eqnarray}\lab{nst}
	\frac{dn }{dt}
	\equiv \left(\frac{\partial}{\partial t}
	+{\rm\bf v} \cdot\mbox{\boldmath$\nabla$}\right)n = - \Gamma_{\rm R}n
	\neq 0  ,
\end{eqnarray}
where
\begin{eqnarray}
	\Gamma_{\rm R}\equiv\frac{1}{\tau_{\rm R}}
	=\frac{2G}{\tau_{\rm GL}} .
\end{eqnarray}

The rate of change of the condensate density ${dn }/{dt}$ is thus described by the relaxation term $\mathcal{R}_{\rm diss} \equiv \Gamma_{\rm R}n$.
From  (\ref{G}) we see that  $\Gamma_{\rm R} = \Gamma_1 + \Gamma_2$,  with
\begin{eqnarray}
	\Gamma_1= 2D_{\rm GL}\left(\frac{m v }{\hbar}\right)^2,
\end{eqnarray}
where the diffusion coefficient $D_{\rm GL} \equiv {\xi^2_{\rm GL}}/{\tau_{\rm GL}}$, and $\Gamma_2$  accounts for non-homogeneities ($\mu_1$ and $n \neq n^{\circ}$) and is related with dissipative processes with life-time usually longer than the GL relaxation time $\tau_{\rm GL}$. Small values of $G$, such that $\tau_{\rm R} = \tau_{\rm GL}/2G \gg \tau_{\rm GL}$,  allow formation of quasi-local equilibrium in the condensate, fast  with respect to longer decay time of the condensate density $n$ ($\tau_{\rm GL} \ll \tau_{\rm R}$). A fast transition to the equilibrium regime occurs at $T_C$; leading to expect
$G \rightarrow 0$ as $T \rightarrow T_C$,  so that  $\tau_{\rm GL} \ll \tau_{\rm R} = \tau_{\rm GL}/2G$, .

The TDGL equation for the normalized wave function $\chi=|\sigma|/|\sigma^{\circ}|\equiv(n /n^{\circ} )^{1/2}$ is obtained from Eq.~(\ref{nst}):
\begin{eqnarray}\lab{nss}
	D_{\rm GL}\mbox{\boldmath$\nabla$}^2\chi-\frac{d\chi}{dt}
	=- \frac{1}{\tau_{\rm GL}}\left[(1-\chi^2)-\xi^2_{\rm GL}\left(\frac{m v }{\hbar}\right)^2\right] \chi~,
\end{eqnarray}
where
$d\chi/dt=\partial\chi/\partial t+{\rm\bf v} \cdot\mbox{\boldmath$\nabla$}\chi$. As explained in the text, for $({1}/{D_{\rm GL}})\, {d\chi}/{dt} = \hbar \ga \, {d\chi}/{dt} \approx 0$,  we get Eq.~(\ref{nssv}) which is recognized to be the vortex equation \cite{DifettiBook,Manka:1990}. For numerical simulations associated to the theoretical analysis see for example \cite{Bettencourt} and refs. quoted therein.

For $\Ga_{\rm R} \approx  \Ga_{1} \equiv 2/{\tau_{\rm E}}$,
and in presence of fast formation of quasi-local equilibrium, one gets
\begin{eqnarray}\lab{tll}
	\frac{\tau_{\rm GL}}{\tau_{\rm E}}
	=\xi^2_{\rm GL}\left(\frac{m v }{\hbar}\right)^2
	\ll 1.
\end{eqnarray}
When the stationary condition $d \chi/d t = 0$ is also met, i.e.  $\partial\chi/\partial t=0$ and
${\rm\bf v} \cdot\mbox{\boldmath$\nabla$}\chi=0$, Eq.~(\ref{nss}) reduces to the stationary GL equation
\begin{eqnarray}\lab{sGL}
	\xi^2_{\rm GL}\mbox{\boldmath$\nabla$}^2\chi+\chi-\chi^3=0 ~.
\end{eqnarray}
%


\end{document}